\documentclass[12pt]{iopart}
\usepackage{pslatex}
\usepackage[T1]{fontenc}
\usepackage[latin1]{inputenc}
\usepackage{graphicx}
\usepackage{hyperref}
\usepackage{bm}
\begin{document}

\title[Testing the performance of a blind burst statistic]{GWDAW 2002}{Testing
the performance of a blind burst statistic}

\author{A~Vicer\'{e}$^{a,b}$\footnote{Corresponding author},
  G~Calamai$^{c,b}$, E~Campagna$^{d,b}$, G~Conforto$^a$, E~Cuoco$^b$,
  P~Dominici$^a$, I~Fiori$^{a,b}$, G~M~Guidi$^{a,b}$, G~Losurdo$^b$,
  F~Martelli$^{a,b}$, M~Mazzoni$^{d,b}$, B~Perniola$^{a,b}$, R~Stanga$^{d,b}$
  and F~Vetrano$^{a,b}$}

\address{$^a$ Istituto di Fisica, Universit\`a di Urbino, Italy.}
\address{$^b$ Istituto Nazionale di Fisica Nucleare, Sez. Firenze/Urbino,
  Italy. }
\address{$^c$ Osservatorio Astrofisico di Arcetri, Firenze, Italy.}
\address{$^d $ Dipartimento di Astronomia, Universit\`a di Firenze, Italy.}

\ead{Andrea Vicer\'e <vicere@fis.uniurb.it>}

\begin{abstract}
In this work we estimate the performance of a method for the detection of
burst events in the data produced by interferometric gravitational wave
detectors. We compute the receiver operating characteristics in the specific
case of a simulated noise having the spectral density expected for Virgo,
using test signals taken from a library of possible waveforms emitted during
the collapse of the core of Type II Supernovae.
\end{abstract}

\submitto{\CQG}
\pacs{04.80.Nn, 07.05.Kf}

\maketitle

\section{Introduction}

Several large scale interferometric instruments are being
commissioned~\cite{GEO,Virgo} or have started science runs~\cite{LIGO,TAMA},
aiming at the detection of gravitational waves (GW) emitted by astrophysical
sources. The interferometers are sensible in a relatively wide frequency band,
which makes it possible to resolve the structure of short bursts of GW, like
those emitted during the Type II supernova explosions, or the merger phase in
the binary black-holes coalescence (see~\cite{arnaud:99} and references
therein).

 However it has been frequently argued~\cite{flanagan:hughes:98,pradier:01}
that the knowledge of the burst waveforms is rather poor, and therefore the
classical Wiener filtering may be not applicable: in this context, it is
advisable to develop methods which do not rely on the theoretical signal
waveform (see~\cite{arnaud:02} for a recent review).
One of us (A.V.) has proposed in~\cite{vicere:01} one such method, which is
claimed to be ``optimal'' under the assumptions that only the duration of the
burst is known, and that the detector noise is Gaussian, but not necessarily
white.

In~\cite{vicere:01} the method has been derived from those assumptions, and
an algorithm has been defined that we will call ``generalized delta filtering''
(GDF), because it reduces itself to a $\delta$-filtering when its ``window
size'' parameter is chosen to be $1$. In that paper no test was made to
assess the actual detection performance of the GDF.

In the present work we have made such a test, using simulated
waveforms~\cite{zwerger:mueller:93} and the expected Virgo spectral noise
density~\cite{virgosensitivity}.

\section{The method in brief}
\label{sec:method}

Referring to~\cite{vicere:01} for details about the method and its derivation,
we just recall that in the single detector case it amounts to apply a few
simple analysis steps:
\begin{enumerate}
\item a $\delta$-filtering step, also called {\em double
  whitening}~\cite{cuoco}, which can
  be implemented in the frequency domain dividing the Fourier transform (FT)
  of the signal by the estimated spectral noise density;
\item estimate the {\em Discrete Karhounen-Lo\`eve} basis for a window of fixed
  size $N$ of the $\delta$-filtered data; we recall that the DKL basis is a
  set of $N$ vectors which ``diagonalize'' the input noise, in the sense that
  coefficients $c_k\equiv\mathbf{x}\cdot{\bm\psi}_k$ of the decomposition of
  data $\mathbf{x}$ over the basis are statistically independent, that is
  $E\left[c_k c_l\right]\propto\delta_{kl}$~\cite{therrien:92};
\item slide the window over the available $\delta$-filtered data and compute
  the statistic
\begin{equation}
\label{eq:statistic}
L=\sum_{k=1}^{N}\frac{1}{\sigma_{k}}\left({\bm\psi}_{k}\cdot\mathbf{x}\right)^{2}
\end{equation}
where $\sigma_k$ and ${\bm\psi}_k$ are the eigenvalues and eigenvectors of the
DKL basis;
\item select candidates by comparing the data against a threshold $\eta$.
\end{enumerate}

The only parameter of the algorithm is the window size $N$, and possibly the
stride used in sliding the window over the $\delta$-filtered data.

\section{The simulation}

\subsection{Input noise}

We have used ``standard'' Virgo noise~\cite{virgosensitivity}, as modeled on
the basis of the expected noise sources~\cite{punturo}, including the thermal
and shot noise in the frequency band $[10,\,10000]$Hz.

To generate noisy data, we started from white Gaussian noise,
``colored'' on the basis of the Virgo sensitivity curve.
\begin{figure}
\begin{center}
\includegraphics[width=0.8\textwidth,clip]{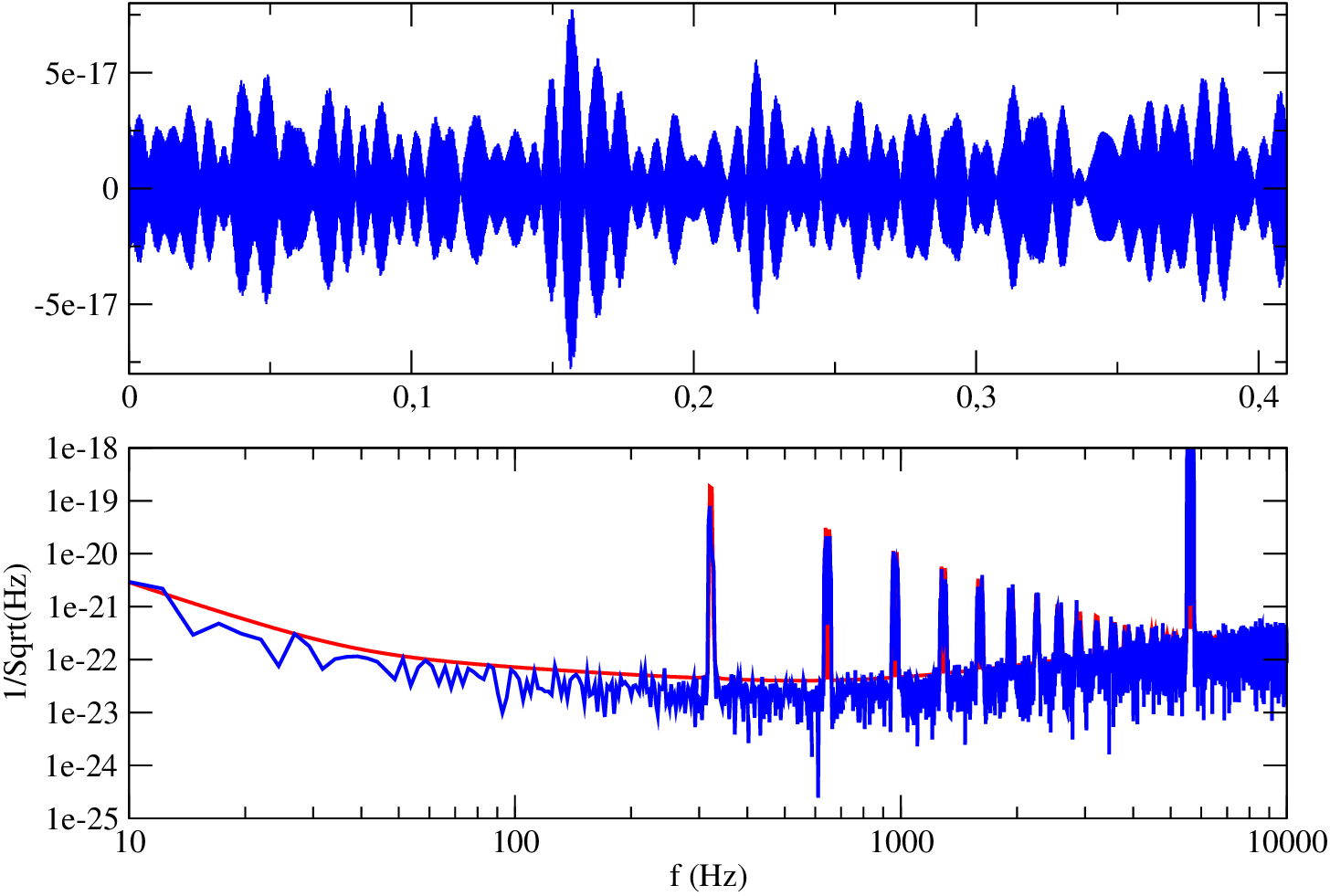}
\caption{\label{fig:virgonoise}In the lower part, the expected spectral noise
  density of the Virgo detector is shown, superimposed to an example of
  spectrum obtained on simulated data. In the upper part, an example of noisy
  output is plotted in the time domain.}
\end{center}
\end{figure}
 We show in Figure~\ref{fig:virgonoise} an example of the simulated noise
obtained using this procedure. Also the resulting spectrum is shown,
superimposed to the ``theoretical'' spectrum. It is worth noting that the
simulated data include the effect of the narrow resonances corresponding to
the suspension ``violin modes'': it is yet not completely clear if these
resonances will be filtered out before applying the detection algorithms, and
we have chosen not to remove them.

\subsection{Simulated signals}

As prototype GW signals, we have used waveforms resulting from
simulations of the core-collapse of Type II supernovae made by Zwerger and
M\"uller~\cite{zwerger:mueller:93}. These waveforms are labeled by the values
of the parameters $A, B, G$ which enter in the polytropic equation of state of
the nuclear matter.

Our method is not optimal in the Wiener sense, and therefore it is crucial to
set the scale of the signals in the simulations. We have chosen to normalize
all the signals to the same {\em intrinsic} signal to noise ratio, defined as
usual by:
\begin{equation}
\textrm{SNR}_{\textrm{intrinsic}}\equiv\sqrt{4\int_{0}^{\infty}\frac{\left|\tilde{s}\left(f\right)\right|^{2}}{S_{n}\left(f\right)}df}
\end{equation}
where $S_n$ is the one-sided spectral density of the noise, and $\tilde{s}$ is
the FT of the signal considered.

In our simulations we have set $\textrm{SNR}_{\textrm{intrinsic}} = 5$ to
compare with the results in~\cite{arnaud:02}.

\subsection{Examples of filtering chain}

It is instructive to have a first exploratory look at the results of a GDF
analysis performed as in Section~\ref{sec:method}.
\begin{figure}
\begin{center}
\includegraphics[width=0.8\textwidth,clip]{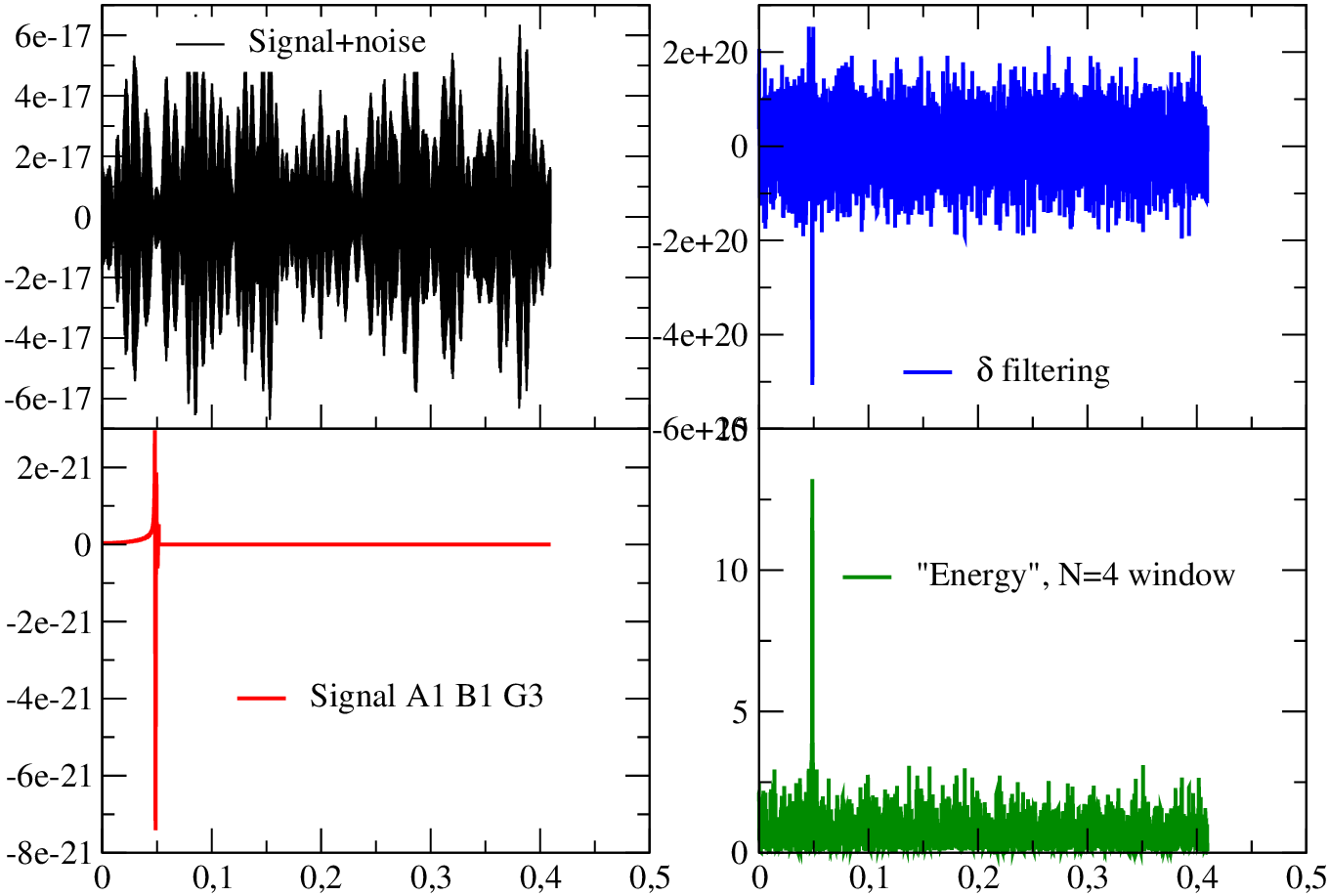}
\caption{\label{fig:example} In anti-clockwise sense: simulated signal+noise,
  signal only, output of a $\delta$-filtering step, output of a GDF analysis
  using an $N=4$ window.}
\end{center}
\end{figure}
In Figure~\ref{fig:example} we show what happens with an $N=4$ window,
choosing a Zwerger-M\"uller (ZM) signal with a rather short duration, and very
much like an instantaneous burst. In this case, the outputs of both the
$\delta$-filtering and the GDF analysis put clearly in evidence the event
occurrence.
\begin{figure}
\begin{center}
\includegraphics[width=0.8\textwidth,clip]{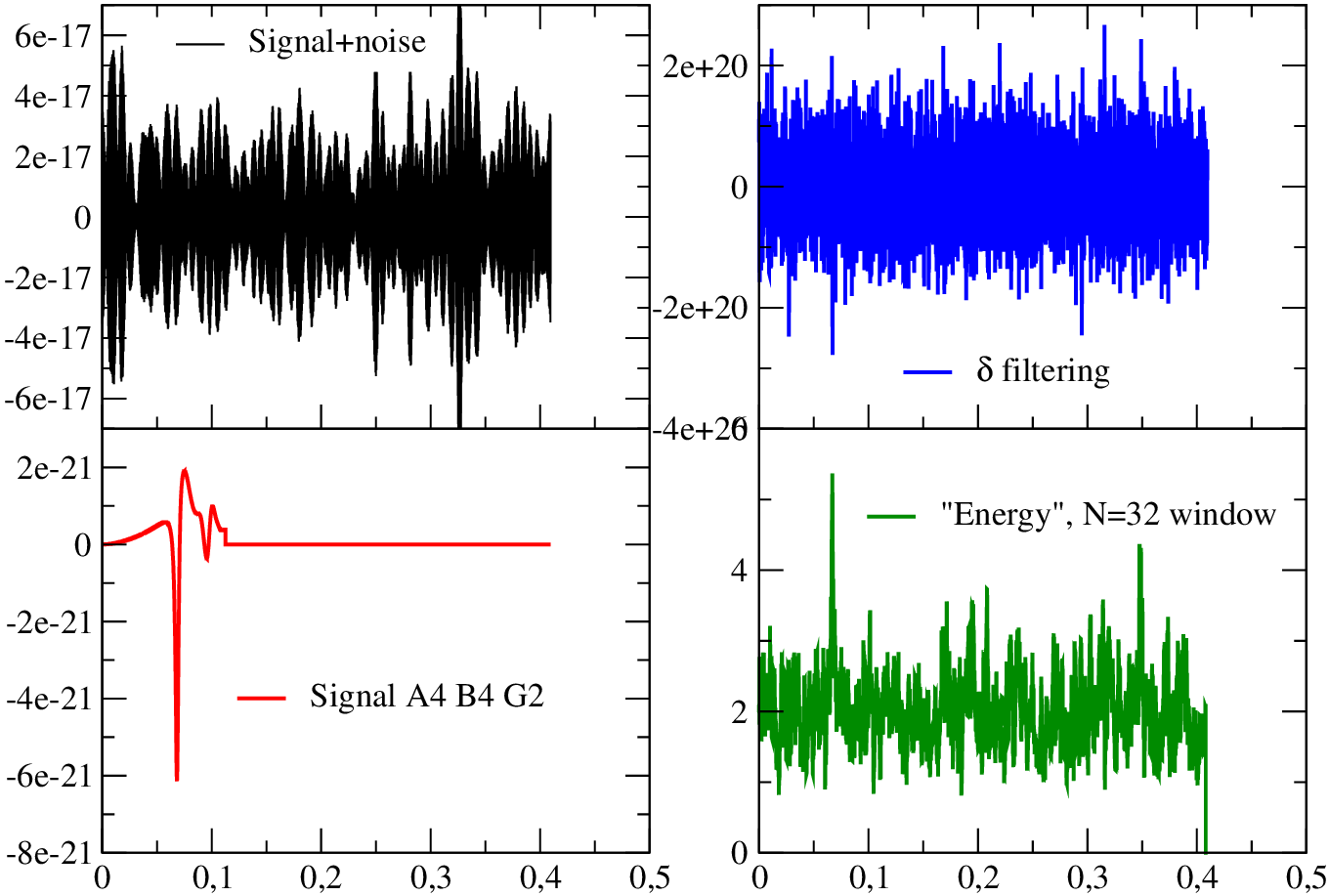}
\caption{\label{fig:example2} In anti-clockwise sense: simulated signal+noise,
  signal only, output of a $\delta$-filtering step, output of a GDF analysis
  using an $N=32$ window.}
\end{center}
\end{figure}
In Figure~\ref{fig:example2} we choose a signal having a more complicated
time-domain evolution: in this case, the $\delta$-filtering is unable to
capture the event, but a GDF analysis using an $N=32$ window appears instead
to put it in evidence, although other false alarms appear in the same time
window, whose significance remains to be assessed.

These examples are just indications that the method works: for a
quantitative understanding we need to estimate the receiver operating
characteristics (ROC).

\section{The ROC estimation}

It is worth recalling that the statistical properties of the GDF detector can
be assessed analytically, because the statistic in Equation~\ref{eq:statistic}
is just a $\chi^2$ with $N$ degrees of freedom.

In presence of a signal, the statistic is a non-central $\chi^2$, and its
distribution is~\cite{vicere:01}
\begin{equation}
\label{eq:distribution}
d\left(L|\textrm{SNR}\right)\textrm{=}\frac{L^{N/2-1}}{s^{N/2}\Gamma\left(\frac{N}{2}\right)}e^{-\frac{1}{2}\left(L+\sqrt{2N}\textrm{SNR}\right)}\,_{0}F_{1}\left(;\frac{N}{2};\frac{\textrm{SNR}\,
    L\sqrt{2N}}{4}\right)
\end{equation}
where $_{0}F_1$ is the hypergeometric function. The distribution
$d\left(L|\textrm{SNR}\right)$ depends only on the $\textrm{SNR}$ of the
signal, as seen by the GDF method. Note that this $\textrm{SNR}$ has nothing
to do with the $\textrm{SNR}_\textrm{intrinsic}$, and it is defined as
customary by the ratio
\begin{equation}
\textrm{SNR}\equiv\frac{E\left[L|H_{1}\right]-E\left[L|H_{0}\right]}{\sqrt{E\left[\left(L-E\left[L|H_{0}\right]\right)^{2}|H_{0}\right]}}
\end{equation}
where $H_{0}$ and $H_{1}$ are the hypotheses of absence or presence of a
signal, respectively. Using this definition it is immediate to show that
\begin{equation}
\textrm{SNR} = \frac{1}{\sqrt{2N}}\sum_{k}\frac{1}{\sigma_{k}}\left({\bm\psi}_{k}\cdot\mathbf{x}\right)^{2}
\end{equation}
for the GDF statistic.

In order to estimate false alarms and detection probabilities, we need
therefore first to compute how much $\textrm{SNR}$ is collected by the GDF
detector(s), for a fixed $\textrm{SNR}_\textrm{intrinsic}$.

\subsection{SNR distribution of the ZM signals}

We have considered the 78 signals in the ZM family~\cite{zwerger:mueller:93},
normalized w.r.t. the theoretical Virgo noise to have
$\textrm{SNR}_\textrm{intrinsic}=5$, and we have computed for each of them the
$\textrm{SNR}$ resulting from a GDF analysis, using different values of the
window size $N$. We stress that there is no point in comparing the values of
$\textrm{SNR}$ for different detection methods or different window sizes $N$:
only the comparison of false alarm and detection probabilities is sensible.
\begin{figure}
\begin{center}
\includegraphics[width=0.8\textwidth,clip]{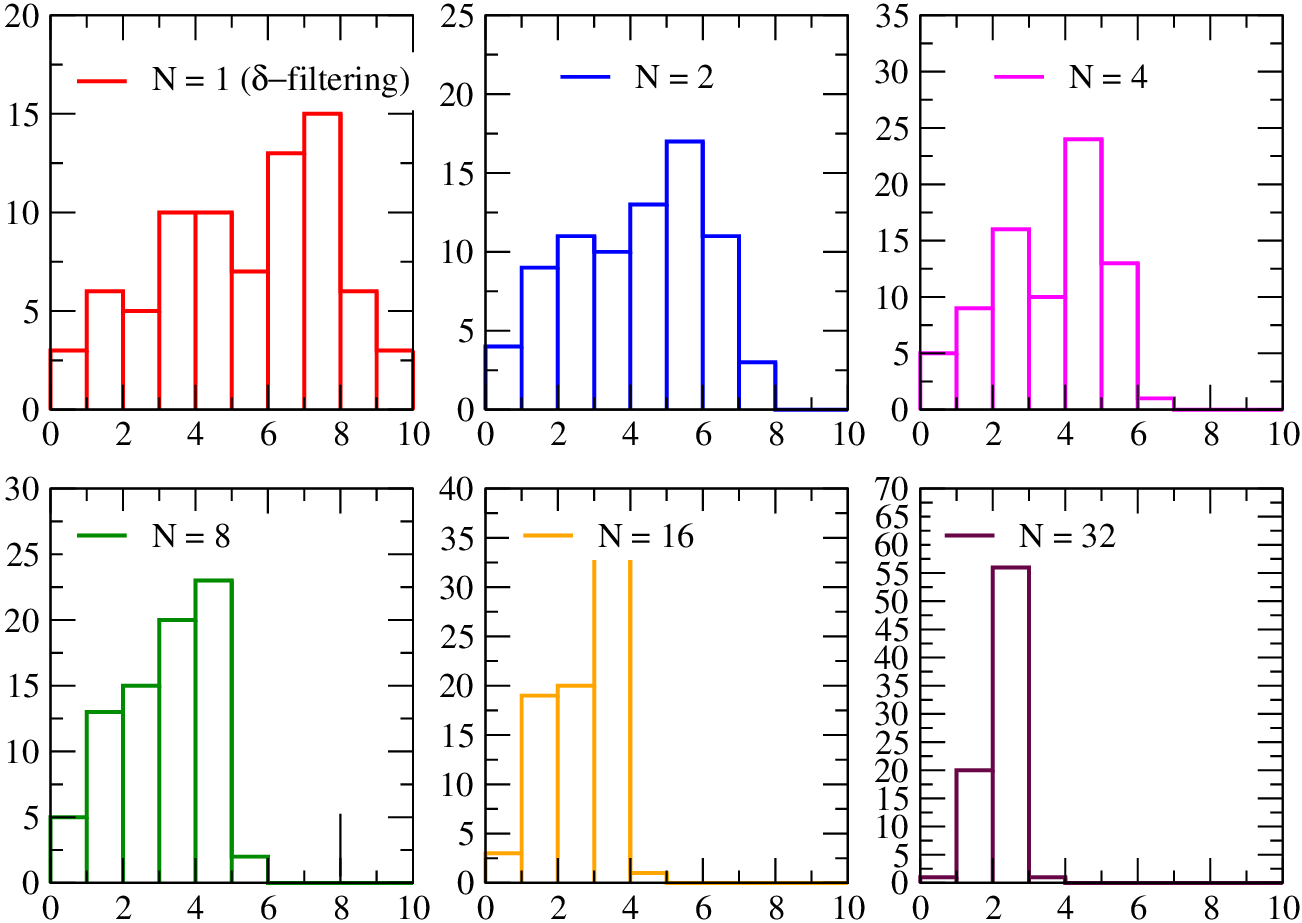}
\caption{\label{fig:snrdistribution} The distribution of $\textrm{SNR}$ for
  different values of the $N$ parameter of the GDF algorithm, based on ZM input
  signals normalized to have $\textrm{SNR}_\textrm{intrinsic}=5$}
\end{center}
\end{figure}
We show in Figure~\ref{fig:snrdistribution} the resulting
$\textrm{SNR}$ distributions.

\subsection{False alarm and detection probabilities}
\label{sec:fadp}

Starting from the distribution in Equation~\ref{eq:distribution} it is
immediate to derive the false alarm probability
\begin{equation}
P_{FA}\left(\eta\right)\equiv\int_{\eta}^{\infty}d\left(L|0\right)dL=\frac{\Gamma\left(\frac{\eta}{2};\frac{N}{2}\right)}{\Gamma\left(\frac{N}{2}\right)}
\end{equation}
while the detection probability $P_{DET}\left(\eta|\textrm{SNR}\right)$
can be evaluated by a numerical integral.

We stress that these $P_{DET}, P_{FA}$ are ``per bin'', namely they do not
take into account the auto-correlation of the filter outputs, which would
require to resample the output or to select, among the outputs above
threshold, only the maxima over a fixed time window, whose length would be
another parameter of the algorithm.

\begin{figure}
\begin{center}
\includegraphics[width=0.8\textwidth,clip]{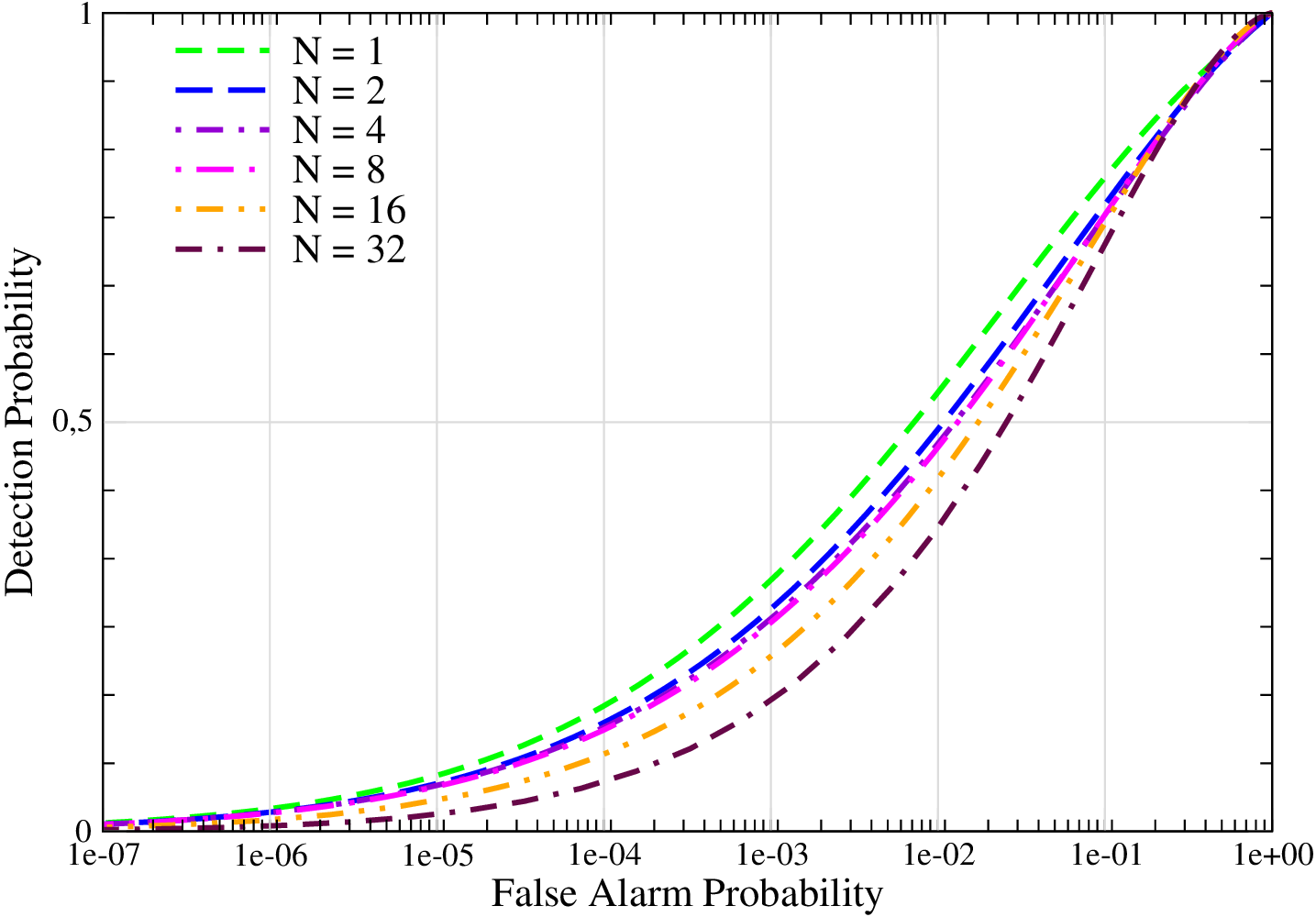}
\caption{\label{fig:rocmean}The ROC curves obtained averaging over the 78
  waveforms of the ZM family (see text), for different window sizes $N$}
\end{center}
\end{figure}

For a given signal, we have varied the threshold $\eta$ to obtain $P_{DET}$
as a function of $P_{FA}$, and for each value of $\eta$ we have averaged the
probabilities over the different ZM waveforms; the resulting plot of $P_{DET}$
against $P_{FA}$ constitutes the average ROC curve, that we show in
Figure~\ref{fig:rocmean} for different values of $N$.
The similarity of the plots may lead to argue that the window size is not very
important. Actually this is just an effect of the averaging, as indicated by
the examples in Figures~\ref{fig:example} and~\ref{fig:example2}, which show
that one should better match the signal length and the window size $N$.
\begin{figure}
\begin{center}
\includegraphics[width=0.8\textwidth,clip]{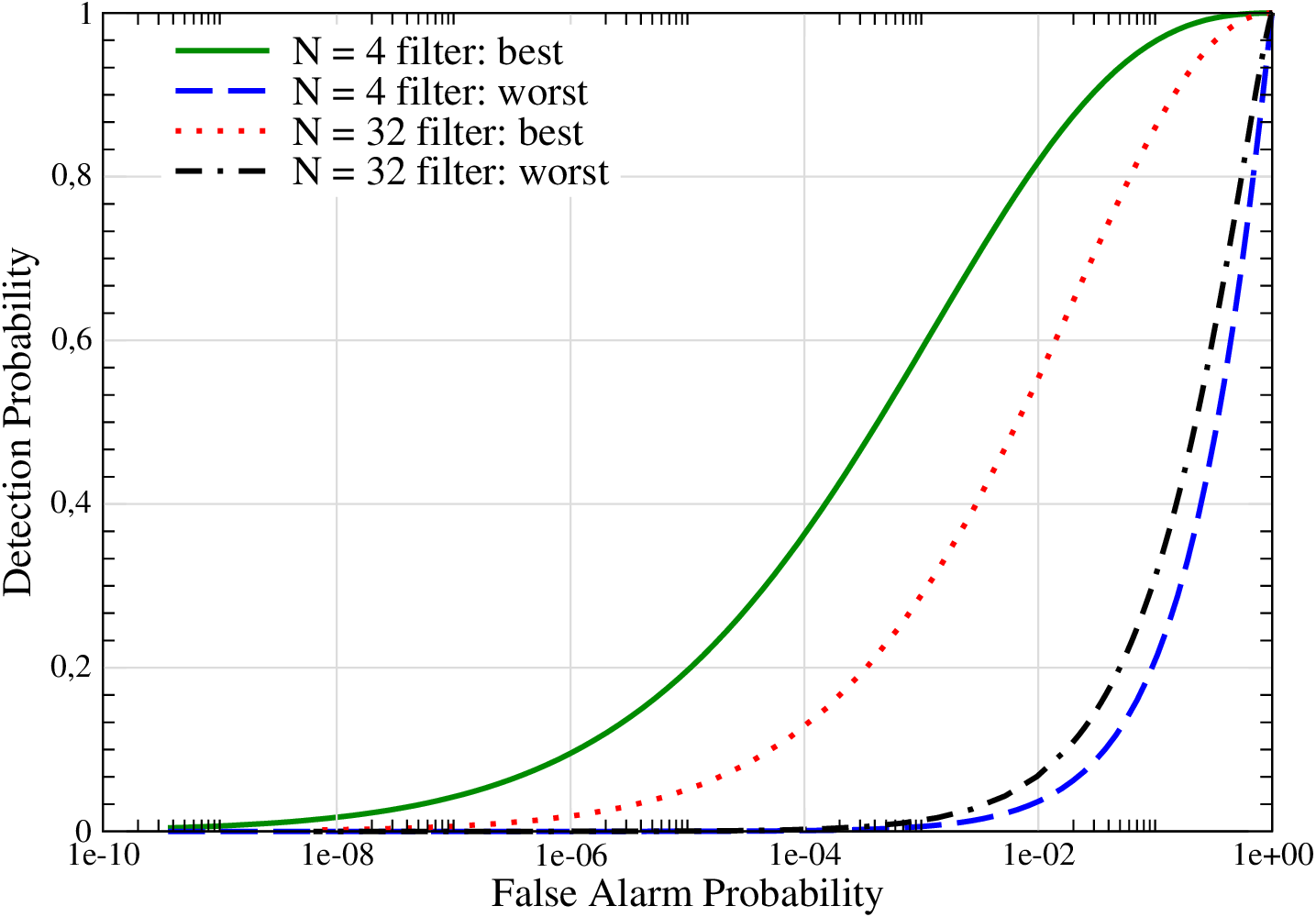}
\caption{\label{fig:rocminmax}The best and worst cases of the ROC curves,
obtained over the different ZM waveforms, for
two different values of the $N$ parameter. Note that the horizontal
scale is different from the one used in Figure~\ref{fig:rocmean}}
\end{center}
\end{figure}
This is confirmed in Figure~\ref{fig:rocminmax}, where we
plot the best and worst ROC curve obtained over the 78 ZM signals, for two
different values (4 and 32) of the $N$ parameter. The large dispersion of the
results is a consequence of the large differences of the SNR recovered by the
GDF filters over the various waveforms, as displayed in
Figure~\ref{fig:snrdistribution}.

\section{Conclusions}

In this paper we have started to assess the detection capabilities of the
``generalized delta filtering'' introduced by one of us in~\cite{vicere:01},
by computing the receiver operating characteristics in the case of Virgo
simulated noise, using signals from the Zwerger-M\"uller family.

The ROC curves we obtain are generally inferior to those
obtained by some other burst detection methods~\cite{arnaud:02}; however we
should mention that in~\cite{arnaud:02} a white Gaussian noise was used, a
difference with our computations whose relevance we are unable to assess.

The results show a wide variation over the $N$ parameter,
and suggest that further optimization is possible: 
one should define a bank of filters with different $N$ values and
understand how to combine the various outputs and how to select
candidates. As discussed in Section~\ref{sec:fadp} one should take
into account the fact that the filter outputs are correlated, and that one may
strongly decrease the false alarm probability by re-sampling the output at a
lower rate, while probably not hampering too much the detection probability.

These studies are beyond the scope of the present work and will be the subject
of a forthcoming paper.

\end{document}